\documentclass[aps,pre,twocolumn,groupedaddress,showpacs,floatfix]{revtex4}
\usepackage{amsmath}
\usepackage{dcolumn}
\usepackage{bm}
\input epsf
\epsfclipon
\begin{document}
\title{Fluctuation-dissipation relations for complex networks}
\author{Agata Fronczak, Piotr Fronczak and Janusz A. Ho\l yst}
\affiliation{Faculty of Physics and Center of Excellence for
Complex Systems Research, Warsaw University of Technology,
Koszykowa 75, PL-00-662 Warsaw, Poland}
\date{\today}

\begin{abstract}
In the paper, we study fluctuations over several ensembles of
maximum-entropy random networks. We derive several
fluctuation-dissipation relations characterizing susceptibilities
of different networks to changes in external fields. In the case
of networks with a given degree sequence, we argue that the
scale-free topologies of real-world networks may arise as a result
of self-organization of real systems into sparse structures with
low susceptibility to random external perturbations. We also show
that the ensembles of networks with noninteracting links (both
uncorrelated and with two-point correlations) are equivalent to
random networks with hidden variables.
\end{abstract}


\maketitle

\section{Introduction}

\par Recently, the statistical properties of real networks
(including biological, social and technological systems) have
attracted a large amount of attention among physicists (see e.g.
\cite{Handbook, Dorogbook, BArev}). It has been realized that
despite functional diversity, most of real web-like systems share
important structural features e.g. small average path length, high
clustering and scale-free degree distribution. A number of network
models have been proposed to embody the fundamental
characteristics. The models can be roughly divided into two
classes: static (homogeneous, equilibrium) and evolving (causal,
nonequilibrium). The second class of causal networks encompasses,
in particular, the famous BA model \cite{BAScience}, whereas
configuration model \cite{Molloy1995,NewPRE2001,Fron2005a} and the
large group of networks with hidden variables
\cite{SodPRE2002,BogPRE2003, Fron2005b} belong to the first class
of static networks. Although very intuitive, the mentioned
representatives of static random networks are {\it not properly}
defined from the point of view of the equilibrium statistical
mechanics. Below we briefly outline what the mentioned lack of
{\it seemliness} means with reference to random networks.

\par To start with, let us concentrate on the phrase {\it random
network}. What does it mean that a network is random? One possible
answer is that there is a large amount of randomness in the
process of network construction. It treats to all the examples of
homogeneous and causal networks quoted in the previous paragraph.
The answer however suffers a few disadvantages among which the
most striking is the issue of quantification of the randomness.
Another answer to the asked question could be that random network
is a member of a statistical {\it ensemble} of networks and the
probability of the occurrence of a given network in random
sampling is proportional to its statistical weight. Without a
doubt, the last treatment directly follows principles of the
equilibrium statistical mechanics.

\par At the moment, a simple example could be the configuration model.
In this model, the total number of nodes is fixed to $N$ and
degrees of all nodes $i=1,2,\dots,N$ create a specific degree
sequence $\{k_i\}$. Until now, nothing has been said about
connections between nodes. As a rule, random graphs with a given
degree sequence are constructed in the following way: i. first,
attach to each node $i$ a number $k_i$ of {\it stubs} (ends of
edges); ii. next, choose pairs of these stubs uniformly at random
and join them together to make complete edges. For sure, such a
procedure represents a large randomness justifying the phrase {\it
random networks}. On the other hand, however, the second among the
mentioned possible meanings of the phrase, treating the resulting
networks as members of the ensemble of graphs with the desired
degree sequence, seems to be more familiar to physicists.

\par As a truth, the concepts of statistical mechanics (including
statistical ensembles, partition functions, averages over ensemble
and so forth) has already been applied to analysis of complex
networks. Although, the majority among the recently submitted
articles still define network models through construction
procedures, there has also been published several interesting
papers on the {\it genuine} statistical mechanics of random
networks (cf. \cite{BurdaPRE2001,
BergPRL2002,Bauer2002,BurdaPRE2003,DorogNuc2003,PallaPRE2004,
Farkas2004,ParkPRE2004,Bogacz2005,Bialas2005}). The general idea
is similar to all the above mentioned papers. Statistical ensemble
of networks is defined by specifying a set of networks
$\mathcal{G}$ which one wants to study (e.g. simple graphs,
digraphs, weighted graphs) and a rule that associates probability
distribution $P(G)$ with these networks $G\in\mathcal{G}$. The
differences between the quoted approaches consist in different
weight assignment strategies.

\par In this contribution we extend the information-theoretic approach
to random networks that was very recently proposed by Park and
Newman \cite{ParkPRE2004} (see also \cite{Bauer2002}). Information
theory \cite{Jayens1,bookCover} provides a criterion for setting
up probability distributions over a given ensemble on the basis of
partial knowledge and leads to a type of statistical inference. It
is the least biased estimate possible on the given information
i.e. it is maximally non-committal with regard to missing
information. Since the procedure consists in entropy maximization
under constraints imposed by the physical conditions of the
ensemble, it is also known as maximum-entropy estimate
\cite{bookKapur}.

\par In this paper, following Park and Newman \cite{ParkPRE2004}, we
use Shannon entropy in order to establish probability distribution
over analyzed networks \cite{footnote1}. Park and Newman have
presented a few exact solutions (in the sense of weighted averages
over ensembles) of specific network models including undirected
networks with a given degree sequence and networks incorporating
arbitrary but independent edge probabilities. Here, we analyze
these exactly solved models from the point of view fluctuations
over ensembles. We discuss several fluctuation-dissipation
relations for the mentioned ensembles. We also show that the
quoted maximum-entropy models are equivalent to random networks
with hidden variables \cite{BogPRE2003}.

\section{General definitions}

\par In this section we review the fundamentals of maximum-entropy random
networks due to Park and Newman \cite{ParkPRE2004}.

\par In order, to correctly define statistical ensemble of networks one
has to start with specifying a set of graphs $\mathcal{G}$ which
one wants to study. In the following, we restrict ourselves to
labelled simple graphs \cite{footnote2} with a fixed number of
nodes $N$. Let us remind that a simple graph has at most one link
between any pair of vertices and it does not contain self-loops
connecting vertices to themselves. Note also that there exists
one-to-one correspondence (isomorphism) between simple graphs and
symmetric matrices of size $N$ with elements $\sigma_{ij}$ equal
either $0$ or $1$.

\par Once the set $\mathcal{G}$ of possible networks has been
established, in the next step one has to decide what kind of
constraints should be imposed on the ensemble. The choice may be,
for example, encouraged by properties of real networks like high
clustering, significant modularity or scale-free degree
distribution. In fact, due to the mentioned isomorphism between
graphs and matrices only such ensembles can be exactly solved
which constraints are simply expressed in terms of adjacency
matrix.

\par Now, suppose that one would like to establish probability
distribution over $\mathcal{G}$ in such a way that the expected
values (i.e. averages over the ensemble) of several observables
$\{x_i(G)\}$, $i=1,2,\dots,r$ were respectively equal to
$\{\langle x_i\rangle\}$. Due to maximum entropy principle the
best choice for probability distribution $P(G)$ is the one that
maximizes the Shannon entropy
\begin{equation}\label{entropy}
S=-\sum_{G}P(G)\ln P(G),
\end{equation}
subject to the constraints
\begin{equation}\label{war1}
\langle x_i\rangle=\sum_{G}x_{i}(G)P(G)
\end{equation}
for $i=1,2,\dots,r$, plus the normalization condition
\begin{equation}\label{norm}
\sum_{G}P(G)=1.
\end{equation}
The Langrangian for the above problem is given by the below
expression
\begin{eqnarray}\label{Lang}
\mathcal{L}&=&-\sum_{G}P(G)\ln P(G)+\alpha(1-\sum_{G}P(G))\\&&+
\sum_{i=1}^{r}\theta_i(\langle x_i\rangle-\sum_{G}x_{i}(G)P(G)),
\end{eqnarray}
where the multipliers $\alpha$ and $\theta_i$ are to be determined
by (\ref{war1}) and (\ref{norm}).

\par Differentiating $\mathcal{L}$ with respect to $P(G)$ and then
equating the result to zero one obtains the desired probability
distribution over the ensemble of graphs with given properties
(\ref{war1})
\begin{equation}\label{PG0}
P(G)=\frac{e^{-H(G)}}{Z},
\end{equation}
where $H(G)$ is the network Hamiltonian
\begin{equation}\label{H0}
H(G)=\sum_{i=1}^{r}\theta_{i}x_{i}(G),
\end{equation}
and $Z$ represents the partition function (normalization constant)
\begin{equation}\label{Z0}
Z=\sum_{G}e^{-H(G)}=e^{\alpha+1}.
\end{equation}

Finally, in order to complete the section devoted to general
considerations it is useful to define the free energy of the
ensemble
\begin{equation}\label{F0}
F=-\ln Z.
\end{equation}
The last quantity is of wide use in the rest of the paper.

\par Now, let us examine the introduced formalism with a few examples.
In the next section, we will analyze fluctuations over the below
presented ensembles.

\subsection{Microcanonical ensemble of random networks}

\par At the beginning, let us study the equivalent of the
microcanonical ensemble for maximum-entropy random networks.
Maximizing Shannon entropy (\ref{entropy}) subject to only
normalization condition (\ref{norm}), i.e. omitting other
constraints (\ref{war1}), one obtains the uniform distribution
over all simple graphs of size $N$
\begin{equation}\label{micPG}
P(G)=\frac{1}{\Omega},
\end{equation}
where $\Omega=2^{N\choose 2}$ represents the total number of the
considered networks i.e. the total number of $0-1$ symmetric
matrices of size $N$. The uniform distribution (\ref{micPG}) means
that each graph in the ensemble have the same weight regardless of
its properties.

\par Since all graphs in the ensemble are equiprobable one can simply
argue that the probability of a graph having $m$ links is given by
\begin{equation}\label{micPm}
P(m)=\frac{{{N\choose 2}\choose m}}{2^{N\choose 2}},
\end{equation}
and respectively
\begin{equation}\label{micsm}
\langle m\rangle=\sum_{m=0}^{N\choose 2}mP(m)=\frac{{N\choose
2}}{2}
\end{equation}
Similarly, the probability of an arbitrary node to have $k$
nearest neighbors equals $P(k)={N-1\choose k}/2^{N-1}$, and the
average connectivity is $\langle k\rangle=(N-1)/2$.

\par In fact, the considered microcanonical ensemble of random networks
is equivalent to the ensemble of classical random graphs with the
connection probability $p=1/2$. Ensembles of classical random
graphs with an arbitrary linkage probability will be considered in
the next subsection.

\subsection{Classical random graphs}

\par Now, let us consider an ensemble of networks with an expected
number of links $\langle m\rangle$ (as stressed before the
ensemble is equivalent to random graphs introduced by
Erd\"os-R\'enyi). The Hamiltonian (\ref{H0}) for this ensemble is
given by
\begin{equation}\label{Her}
H(G)=\theta m(G),
\end{equation}
where $\theta$ represents a field or an inverse temperature whose
value is fixed and depends only on $\langle m\rangle$. Park and
Newman \cite{ParkPRE2004} have shown that the partition function
(\ref{Z0}) for the ensemble is equal to
\begin{equation}\label{Zer}
Z=(1+e^{-\theta})^{N\choose 2},
\end{equation}
and respectively the free energy (\ref{F0}) can be written as
\begin{equation}\label{Fer}
F=-\ln Z=-{N\choose 2}\ln (1+e^{-\theta}).
\end{equation}

\par Having probability distribution (\ref{PG0}) over the ensemble one
can, for example, find the relation between the average number of
links $\langle m\rangle$ and $\theta$
\begin{equation}\label{mer}
\langle m\rangle=\frac{\partial F}{\partial\theta}
=\frac{{N\choose 2}}{e^\theta+1}.
\end{equation}
Now, since $\theta$ is fixed one can reexpress the last formula in
terms of the linkage probability $p$ that is known from the theory
of classical random graphs
\begin{equation}\label{mer1}
\langle m\rangle={N\choose 2}p,
\end{equation}
where
\begin{equation}\label{per}
p=\frac{1}{e^\theta+1}.
\end{equation}

\par Finally, let us point out that in the limit of very small fields
$\theta\rightarrow 0$ (high temperatures) the ensemble of random
networks with an expected number of links is equivalent to the
microcanonical ensemble of random networks (\ref{micPG})
introduced in the previous subsection.

\subsection{Networks with a given degree sequence}

\par At the moment, suppose that one would like to deal with random
networks with an expected degree sequence
\begin{equation}\label{Pk_seq}
\{\langle k_i\rangle\}\;\;\;\;\;\mbox{for}\;\;\;\;\;
i=1,2,\dots,N.
\end{equation}
In this case, the network Hamiltonian is given by
\begin{equation}\label{PkH}
H(G)=\sum_{i=1}^{N}\theta_{i}k_{i}(G),
\end{equation}
where the multipliers $\theta_i$ represent a kind of potential
assigned to each node and they only depend on the expected degrees
$\langle k_i\rangle$ (see Eqs. (\ref{Pkki}) and (\ref{Pkki1})).
The partition function for the considered ensemble can be written
as \cite{ParkPRE2004}
\begin{equation}\label{PkZ}
Z=\prod_{i<j}(1+e^{-(\theta_i+\theta_j)}),
\end{equation}
and the free energy is
\begin{equation}\label{PkF}
F=-\sum_{i<j}\ln(1+e^{-(\theta_i+\theta_j)}),
\end{equation}

\par Performing weighted averages over the ensemble one can easily
prove the below identities: the first one describing the
connection probability between two nodes $i$ and $j$
\begin{equation}\label{Pkpij}
p_{ij}=\frac{1}{e^{(\theta_i+\theta_j)}+1},
\end{equation}
and the second identity for the average connectivity of a node
characterized by the local field $\theta_i$
\begin{equation}\label{Pkki}
\langle k_i\rangle(\theta_i)=\frac{\partial F}{\partial\theta_i}
=\sum_{j=1}^{N}\frac{1}{e^{(\theta_i+\theta_j)}+1}=\sum_{j=1}^{N}p_{ij}.
\end{equation}
The both expressions show the reverse relation between the two
parameters (i.e. $\langle k_i\rangle$ and $\theta_i$)
characterizing each node. The relation consist in the statement:
small degrees correspond to large multipliers and vice versa,
nodes with a large number of connections possess small
multipliers.

\par Park and Newman \cite{ParkPRE2004} have pointed out that
instead of studying networks with an expected degree sequence
(\ref{Pk_seq}) one can deal with networks characterized by an
expected degree distribution $P(\langle k_i\rangle)$
\cite{footnote3}. The authors have argued that one can produce any
degree distribution by a judicious choice of the distribution of
multipliers $\rho(\theta_i)$. In fact, due to (\ref{Pkki}),
$\rho(\theta_i)$ resulting in the desired $P(\langle k_i\rangle)$
can be determined from the below expression
\begin{equation}\label{PkRt}
\rho(\theta_i)=P(\langle k_i\rangle)\left|\frac{d\langle
k_i\rangle}{d\theta_i}\right|,
\end{equation}
where $\langle k_i\rangle(\theta_i)$ is given by (\ref{Pkki}).
There are, however, a few subtleties related to the transition
between the sequence $\{\langle k_1\rangle,\langle
k_2\rangle\,\dots,\langle k_N\rangle\}$ and $P(\langle
k_i\rangle)$. First, the Eq. (\ref{PkRt}) defines $\rho(\theta_i)$
as an implicit function which, except for a very few cases, can
not be explicitly calculated. Second, performing such a transition
one has to keep in mind that our phase space consists of labelled
graphs in which every node $i=1,2,\dots,N$ has assigned its own
multiplier $\theta_i$ (i.e. is distinguishable). Using
$\rho(\theta_i)$ makes an impression that the nodes lose their
identities. In such a case, there exists a threat of widening the
original phase space.

\par To proceed further, let us consider {\it sparse} networks. In
the case, connection probabilities (\ref{Pkpij}) factorize
\begin{equation}\label{Pkpij1}
p_{ij}\simeq e^{-(\theta_i+\theta_j)}=\frac{\langle k_i
\rangle\langle k_j\rangle}{\langle k\rangle N},
\end{equation}
where
\begin{equation}\label{Pkki1}
\langle k_i\rangle(\theta_i) \simeq e^{-\theta_i}\sqrt{\langle
k\rangle N}.
\end{equation}
As shown in \cite{BogPRE2003,Fron2005b} such ensembles are
equivalent to uncorrelated networks. The relation (\ref{Pkki1})
between expected degrees and its multipliers makes the ensembles
very simple for both Monte Carlo simulations and analytical
treatment. In particular, the distribution of multipliers
(\ref{PkRt}) corresponding to $P(\langle k_i\rangle)$ is simply
\begin{equation}\label{PkRt1}
\rho(\theta_i)=\langle k_i\rangle P(\langle k_i\rangle),
\end{equation}
where $\langle k_i\rangle$ is given by (\ref{Pkki1}).

\par There exist, however, some side effects of the approximation.
First, since the connection probability $p_{ij}\leq 1$,
(\ref{Pkpij1}), thus the assumption of sparse networks is only
valid for networks with non-negative Lagrange multipliers (i.e.
$\theta_i,\theta_j\geq 0$). The restriction causes failure of the
approach in the case of scale-free networks $P(k)\sim k^{-\gamma}$
with $2<\gamma<3$. To existence of hubs $k_{max}\sim
N^{1/(\gamma-1)}$ \cite{footnote4}, i.e. nodes with negative
multipliers (see comment after Eq. (\ref{Pkki})), spontaneously
develop degree correlations \cite{MaslovPhysA2004,ParkPRE2003}. It
was argued \cite{BurdaPRE2003,BogEPJ2004,CatanPRE2005} that one
can omit the correlations by applying the so-called structural
cutoff i.e. forcing the largest degree to scale as
$k_{max}\sim\sqrt{N}$. At the moment, let us stress that the
structural cutoff in uncorrelated networks naturally emerges form
the Eq. (\ref{Pkki1}) when $\theta_i\rightarrow 0$.

\subsection{Networks with two-point correlations}

\par In order to study random networks with two-point correlations one
may consider a class of Hamiltonians (\ref{H0}) constructed on the
basis of an expected linkage probability
\begin{equation}\label{corH}
H(G)=\sum_{i<j} \Theta_{ij}\sigma_{ij}(G).
\end{equation}
In the last expression $\sigma_{ij}(G)$ is an element of the
adjacency matrix representing the graph $G$ and $\Theta_{ij}$
characterizes field coupled to the hypothetical link $\{i,j\}$.
The partition function and the free energy for the ensemble are
given by
\begin{equation}\label{corZF}
Z=\prod_{i<j}(1+e^{-\Theta_{ij}}),\;\;\;\;\;
F=-\sum_{i<j}\ln(1+e^{-\Theta_{ij}}).
\end{equation}
Comparing (\ref{PkZ}) and (\ref{corZF}) one can see that the
previous ensemble of networks with an expected degree sequence is
a special case (for $\Theta_{ij}=\theta_i+\theta_j$) of networks
with arbitrary two-point correlations. Similarly to (\ref{Pkpij})
one can also find that
\begin{equation}\label{corpij}
p_{ij}=\langle\sigma_{ij}\rangle=\frac{\partial
F}{\partial\Theta_{ij}}= \frac{1}{e^{\Theta_{ij}}+1}.
\end{equation}

\section{Fluctuations and responses}

\par In classical thermodynamics, fields interacting with a system
have conjugate variables which represent the response of the
system to perturbation of the corresponding field. For example,
the response of a gas to a change in volume is a change in
pressure. The pressure $p$ is the conjugate variable to the volume
$V$. Similarly, the magnetization $M$ of a magnet changes in
response to the applied field $B$. The mentioned relations are
produced by terms in the Hamiltonian of the form $YX$, where $Y$
is a field and $X$ is the conjugate variable to which it couples.
Note, that the above also holds for maximum-entropy random
networks being the subject of the paper, where (see Eq.
(\ref{H0}))
\begin{equation}\label{H00}
H(G)=\sum_{i=1}^{r}\theta_ix_i(G).
\end{equation}

\par Taking advantage of (\ref{PG0})-(\ref{F0}), expectation
values $\langle x_{i}\rangle$ of observables $x_{i}$ can be
calculated as first derivatives of the free energy with the
appropriate field $\theta_i$ (cf.
(\ref{mer}),(\ref{Pkki}),(\ref{corpij}))
\begin{equation}\label{sxi0}
\langle x_i\rangle=\sum_{G}x_i(G)P(G)=\frac{\partial F}{\partial
\theta_i}.
\end{equation}
Similarly, second derivatives of the free energy $F$ give the mean
square fluctuations of the variables
\begin{equation}\label{varxi0}
\langle x_{i}^2\rangle-\langle x_i\rangle^2=-\frac{\partial^2
F}{\partial\theta_i^2}.
\end{equation}
Now, inserting (\ref{sxi0}) into (\ref{varxi0}) one obtains a very
important result
\begin{equation}\label{fdt0}
\langle x_{i}^2\rangle-\langle x_i\rangle^2=-\frac{\partial
\langle x_i\rangle}{\partial\theta_i}=\chi_i^{(\theta)},
\end{equation}
that is known as the {\it fluctuation-dissipation theorem} (FDT).
The theorem states that fluctuations in an observable $x_{i}$ are
proportional to the susceptibility $\chi_i^{(\theta)}$ of the
observable to its conjugate field $\theta_i$. Let us remind that
the susceptibility $\chi_i^{(\theta)}$ measures the strength of
the response of $x_{i}$ to changes in $\theta_i$. In reality, due
to practical purposes, it is often simpler to analyze the
susceptibility $\chi_i^{(\phi)}$ to other field $\phi_i$ that
directly depends on $\theta_i$ (\ref{fdt0})
\begin{equation}\label{fdt0a}
-\frac{\partial \langle
x_i\rangle}{\partial\theta_i}=-\frac{\partial \langle
x_i\rangle}{\partial\phi_i}\frac{\partial\phi_i}
{\partial\theta_i}=\frac{\partial\phi_i}{\partial\theta_i}\chi_i^{(\phi)},
\end{equation}
where $\partial\phi_i/\partial\theta_i$ is the transitional
derivative.

Probably the best known example of the theorem (\ref{fdt0}) is the
one arising from fluctuations of energy in the canonical ensemble
\begin{equation}\label{fdtE}
\langle E^2\rangle-\langle E\rangle^2=-\frac{\partial
E}{\partial\beta}=kT^2C_{V},
\end{equation}
where $C_{V}=\partial\langle E\rangle/\partial T$ is the heat
capacity (or thermal susceptibility), whereas
$kT^2=\partial\beta/\partial T$ is the respective transitional
derivative. Another example relates fluctuations in the
magnetization to the magnetic susceptibility
\begin{equation}\label{fdtM}
\langle M^2\rangle-\langle M\rangle^2=
\frac{1}{\beta}\frac{\partial M}{\partial B}=
\frac{1}{\beta}\chi^{(B)}.
\end{equation}

The fluctuation-dissipation theorems (\ref{fdt0})-(\ref{fdtM}) are
interesting for a number of reasons. First, they join both
microscopic description (left-hand side) and macroscopic,
properties (right-hand side) of the considered systems. Second,
they relate the actual state (fluctuations) of the systems to
their future behavior (response). Third, due to FDT phase
transitions certified by singularities in susceptibilities can
also be reported by large scale fluctuations.

Extending the idea of the susceptibility one can consider what
happens with a variable $x_i$ when one changes the value of a
field $\theta_j$. To study the problem one can define a
generalized susceptibility $\chi_{ij}^{(\theta)}$ which is a
measure of the response of $\langle x_i\rangle$ to the variation
of the field $\theta_j$
\begin{equation}
\chi_{ij}^{(\theta)}=-\frac{\partial\langle
x_{i}\rangle}{\partial\theta_j}
\end{equation}
Again, the susceptibility $\chi_{ij}^{(\theta)}$ is a second
derivative of the free energy
\begin{equation}
\chi_{ij}^{(\theta)}=-\frac{\partial^2
F}{\partial\theta_i\partial\theta_j} =\langle
x_ix_j\rangle-\langle x_i\rangle\langle x_j\rangle.
\end{equation}
The issue of generalized susceptibilities is of special interest
in lattice systems (\ref{H00}), where the variables $x_{i}$ for
$i=1,2,\dots,r$ may represent to the same observable but measured
in different spatial points. Then, susceptibility
$\chi_{ij}^{(\theta)}$ is just the two-point correlation function
between sites $i$ and $j$.

In the following, we will concentrate on fluctuations over a few
selected ensembles of random networks.

\subsection{Classical random graphs}

At the beginning, let us consider the ensemble of classical random
graphs. By definition, the average number of links $\langle
m\rangle$ is fixed in the ensemble. As stressed at the beginning
of the section, there exist, however, fluctuations around the
average. In fact, the probability of a graph $G$ with $m$ links is
given by
\begin{equation}\label{er1PG}
P(G)=\frac{e^{-\theta m(G)}}{Z}=p^m(1-p)^{{N\choose 2}-m}.
\end{equation}
The variance of the above distribution calculated from
(\ref{fdt0}) is very similar to (\ref{fdtE})
\begin{eqnarray}\label{erfdtm}
\langle m^2\rangle-\langle m\rangle^2&=&-\frac{\partial\langle
m\rangle}{\partial\theta}\\&=&-\frac{\partial\langle
m\rangle}{\partial p}\frac{\partial
p}{\partial\theta}=p(1-p)C_{m}\nonumber,
\end{eqnarray}
where $C_{m}=\partial\langle m\rangle/\partial p={N\choose 2}$ is
the link capacity in classical random graphs. Note that for a
given network size $N$ the link capacity does not depend on the
linkage probability $C_m(p)=const$ (classical ideal gases reveal
the analogous behavior $C_{V}(T)=const$).

\subsection{Networks with a given degree sequence}

Now, let us continue with random networks characterized by an
expected degree sequence (\ref{Pk_seq}).

Taken advantage of (\ref{Pkpij}) and (\ref{Pkki}),
fluctuation-dissipation relations (\ref{fdt0}) for the ensemble
may be written in the following form
\begin{eqnarray}\label{Pk_fdt1}
\chi_i^{(\theta)}=-\frac{\partial\langle
k_i\rangle}{\partial\theta_i}&=&\langle k_i^2\rangle-\langle
k_i\rangle^2\\&=&\sum_{j}p_{ij}(1-p_{ij})=\langle
k_i\rangle-\sum_{j}p_{ij}^2\nonumber.
\end{eqnarray}
At the moment, before delving into the discussion of the last
expression, let us note that the susceptibility
$\chi_{i}^{(\theta)}$ is also given by the below formula
\begin{eqnarray}
\chi_{i}^{(\theta)}=-\sum_{j}\frac{\partial\langle k_i\rangle}
{\partial p_{ij}}\frac{\partial
p_{ij}}{\partial\theta_i}=-\sum_{j}\frac{\partial
p_{ij}}{\partial\theta_i}C_{ij},
\end{eqnarray}
where $C_{ij}=\partial\langle k_i\rangle/\partial p_{ij}$
represents the link capacity. Here, since $C_{ij}=1$ the
fluctuations in nodes degrees result only form the transitional
derivative $\partial p_{ij}/\partial\theta_i$.

The importance of the two above identities lies in the fact that
from the fluctuations over degrees of nodes characterized by the
same local field $\theta_i$, one can deduce on the future behavior
of the nodes in the face of possible changes in $\theta_i$. Large
(small) fluctuations correspond to high (low) local susceptibility
$\chi_i^{(\theta)}$.

Now, let us note, that in the case of small degrees (see also the
assumption of sparse networks (\ref{Pkki1})), the
fluctuation-dissipation relation (\ref{Pk_fdt1}) gets a simplified
form
\begin{equation}\label{Pk_fdt2} \langle k_i^2\rangle-\langle
k_i\rangle^2\simeq\langle k_i\rangle,
\end{equation}
indicating the Poissonian fluctuations. Since however, in the case
of sparse homogeneous networks one can omit the last term in
(\ref{Pk_fdt1}), in sparse scale-free networks with the
characteristic exponent $2<\gamma<3$, the mentioned term is
dominated by hubs and the nodes susceptibilities
$\chi_i^{(\theta)}$ are much smaller than their expected degrees
$\langle k_i\rangle$. The total network susceptibility decreases
making the system resistant against random changes in the
landscape of multipliers and simultaneously susceptible to
behavior of supernodes \cite{HavPRL2000,HavPRL2001}.

\begin{figure} \epsfxsize=7.5cm \epsfbox{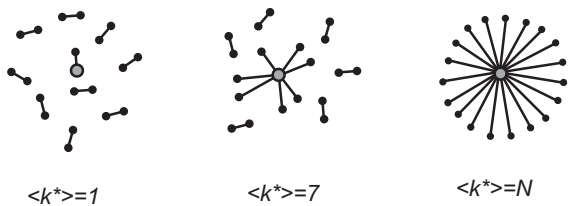}
\caption{Schematic representation of networks possessing $N=19$
nodes with one nearest neighbor $\langle k\rangle=1$ and the
supernode (the gray one) with the tunable connectivity $\langle
k^*\rangle=1,2, \dots,N$.}\label{fig1}
\end{figure}

In order to establish the better understanding of the statement
included in the last paragraph, let us consider a trivial network
consisting of $N$ nodes with expected degrees $\langle k\rangle=1$
and one {\it supernode} with the tunable desired degree $\langle
k^{*}\rangle=1,2,\dots,N$ (in the sequel, the parameters denoted
by the star apply to the supenode). Adjusting the degree of the
supernode makes possible to smoothly pass between regular graphs
(for $\langle k^{*}\rangle=1$) and star networks (for $\langle
k^{*}\rangle=N$) (see Fig. \ref{fig1}). The transition enables the
understanding of what the reduced network susceptibility consists
in.

\begin{figure} \epsfxsize=7.8cm \epsfbox{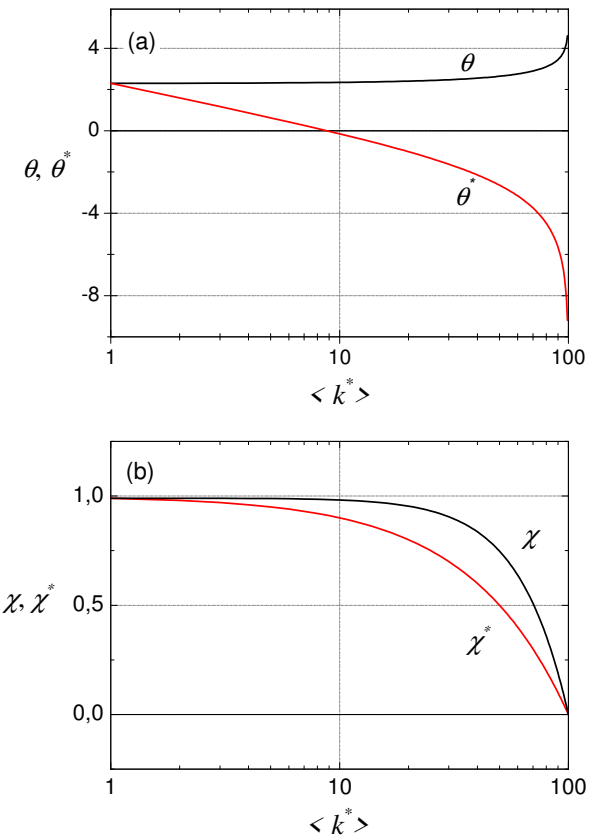}
\caption{(a) Lagrange multipliers $\theta$ and $\theta^*$ as a
function of $\langle k^*\rangle$, (b) relative susceptibilities
$\chi$ and $\chi^{*}$ as a function of $\langle k^*\rangle$. Here,
we have assumed $N=100$.} \label{fig2}
\end{figure}

First, let us find the Lagrange multipliers $\theta$ and
$\theta^{*}$ corresponding to the nodes of the considered
ensemble. Taking advantage of (\ref{Pkki}) one can show, that the
parameters fulfill the set of equations
\begin{displaymath}
\left\{
\begin{array}{ccl}1&=&
\dfrac{N-1}{e^{2\theta}+1}+\dfrac{1}{e^{\theta+\theta^{*}}+1}\\\langle
k^{*}\rangle&=&\dfrac{N}{e^{\theta+\theta^{*}}+1}.
\end{array}
\right.
\end{displaymath}
Solving the above equations for $\theta$ and $\theta^*$ one gets
(see Fig. \ref{fig2}a)
\begin{displaymath}
\left\{
\begin{array}{ccl}
\theta&=&\dfrac{1}{2}\ln\left[\dfrac{N^2-2N+\langle k^{*}\rangle}
{N-\langle k^{*}\rangle}\right]\\\theta^{*}&=&\ln\left[
\dfrac{N-\langle k^{*}\rangle}{\langle k^{*}\rangle}\right]-
\theta\nonumber.
\end{array}
\right.
\end{displaymath}
Next, inserting the multipliers into (\ref{Pkpij}) and then taking
advantage of (\ref{Pk_fdt1}) one obtains the susceptibilities of
expected degrees due to changes in the nodes intensive parameters.
The relative susceptibilities are respectively given by
\begin{eqnarray}\label{Pk_fdtex1}
\chi=-\frac{1}{\langle k\rangle}\frac{\partial\langle k
\rangle}{\partial\theta}=1-\frac{(N-\langle
k^*\rangle)^2}{N^2(N-1)}-\frac{\langle k^*\rangle^2}{N^2}
\end{eqnarray}
for the bulk of nodes, and
\begin{eqnarray}\label{Pk_fdtex2}
\chi^{*}=-\frac{1}{\langle k^*\rangle}\frac{\partial\langle k^*
\rangle}{\partial\theta^*}=1-\frac{\langle k^*\rangle}{N}
\end{eqnarray}
for the supernode.

The behaviour of susceptibilities (\ref{Pk_fdtex1}) and
(\ref{Pk_fdtex2}) is depicted at the Fig. \ref{fig2}b. One can see
that the susceptibilities decrease to $0$ when the expected degree
of the hub $\langle k^*\rangle$ approaches $N$. In the region of
the vanishing susceptibilities, the small changes in the fields
$\theta$ and $\theta^*$ poorly affect the topological features
(i.e. $\langle k^*\rangle$) of the considered networks (c.f. Fig.
\ref{fig2}a). The last statement supports our previous claim of
the resilience of such networks against random external
perturbations. The large $\langle k^*\rangle$ makes that the
supernode accumulates most of the links present in the system and
causes that there exist relatively small number of network
realizations which both: i. fulfill physical constraints of the
ensemble and ii. possess significant statistical weights. For
example, only one such a realization exists in the limiting case
of the star network with $k^*=N$.

\section{Equivalence of maximum entropy networks and networks with hidden variables}

In the section, we continue the thread of Poissonian fluctuations
(\ref{Pk_fdt2}).

Random networks with hidden variables are simply defined through
the construction procedure that consists of only two steps:
\begin{itemize}
\item[i.] first, prepare $N$ nodes and assign them hidden
variables independently drawn from the probability distribution
$R(h)$,
\item[ii.] next, each pair of nodes $\{i,j\}$ link with a
probability $r_{ij}$.
\end{itemize}
One can show, that the uncorrelated networks with hidden variables
arise from the factorized connection probability
\begin{equation}\label{rij}
r_{ij}=\frac{h_{i}h_{j}}{\langle h\rangle N},
\end{equation}
whereas networks with two-point correlations require more
sophisticated expressions for $r_{ij}$.

Even comparing the above short review to our previous results on
sparse networks with an expected degree sequence (cf. Eqs.
(\ref{Pkpij1}) and (\ref{rij})), allows one to deduce on the
equivalence of the two approaches. In the course of the section,
we will argue that the claimed equivalence also holds for networks
with two-point correlations. We will prove it by recovering the
so-called Poissonian propagators characterizing both correlated
and uncorrelated sparse networks with hidden variables
\cite{BogPRE2003}.

\subsection{Networks with a given degree sequence}

At the moment, it is clear that although the expected degree of
the node $i$ is $\langle k_i\rangle$, but due to ensemble
fluctuations its actual degree $k_i$ changes from network to
network. Our aim is to find the so-called propagator
$P(k_i/\theta_i)$ i.e. the degree distribution of the node given
that it is characterized by the multiplier $\theta_i$.

At the beginning, let us reformulate the probability of a graph
$G$ in the ensemble
\begin{equation}
P(G)=\frac{e^{-H(G)}}{Z},
\end{equation}
where $H(G)$ and $Z$ are respectively the graph Hamiltonian
(\ref{PkH}) and the partition function (\ref{PkZ}). Taking
advantage of the connection probability $p_{ij}$ (\ref{Pkpij}),
$P(G)$ can be written in a similar form as in the case of
classical random graphs (\ref{er1PG})
\begin{equation}\label{Pk1PG}
P(G)=\prod_{i<j}\Phi(i,j),
\end{equation}
where
\begin{equation}
\Phi(i,j)=p_{ij}^{\;\;\sigma_{ij}}(1-p_{ij})^{(1-{\sigma_{ij}})}
\end{equation}
whereas $\sigma_{ij}$ are elements of the adjacency matrix
describing $G$ and they are equal to either $1$ or $0$ depending
on whether $i$ and $j$ are connected or not.

In the following, without the loss of generality, we will
concentrate on the node $i=1$. Having $P(G)$ one can calculate the
probability $P(\{\sigma_{1j}\})$ of the node to have a given
linkage profile $\{\sigma_{1j}\}$ (e.g. \{0,0,0,1,0,1,\dots,0\})
\begin{eqnarray}
P(\{\sigma_{1j}\})&=&\sum_{G^*}P(G)\nonumber\\&=&\prod_{j}\Phi(1,j)
\prod_{2\leq i<j}\;\sum_{\sigma_{ij}=0}^{1}\Phi(i,j)\nonumber
\\&=&\prod_{j}\Phi(1,j)\label{Pk2PG},
\end{eqnarray}
where the first sum runs over all networks $G^{*}$ with the fixed
sequence $\{\sigma_{1j}\}$ (i.e. the fixed neighborhood of the
node $i=1$). Now, in order to obtain $P(k_1/\theta_1)$ one has to
sum the probabilities (\ref{Pk2PG}) over different sequences
$\{\sigma_{1j}\}^*$ representing the same degree
$k_1=\sum_{j}\sigma_{1j}$
\begin{eqnarray}\label{prop0}
P(k_1/\theta_1)=\sum_{\{\sigma_{1j}\}^*}P(\{\sigma_{1j}\}).
\end{eqnarray}

Up to this point, the derivation of $P(k_1/\theta_1)$ has been
exact. Now, before proceeding with approximations let us test the
formula (\ref{prop0}) against the simplest ensemble i.e. networks
with an expected homogenous degree sequence. In the ensemble, all
nodes have the same desired degree and also
$\forall_{i=1}^{N}\;\theta_i=\theta$. It is easy to check that the
degree distribution of an arbitrary node (\ref{prop0}) is given by
\begin{equation}\label{proper}
P(k/\theta)={N-1\choose k}p^{k}(1-p)^{N-1-k},
\end{equation}
where the binomial factor in the front of the expression arises
from the fact that there exist ${N-1\choose k}$ different
connection profiles corresponding to degree $k$ and
$p=(e^{2\theta}+1)^{-1}$ (\ref{Pkpij}) (please do not confuse it
with (\ref{per}), where $\theta$ has a different meaning !). One
should not be surprised with the last result. If it is not
obvious, let us stress that the ensemble of networks with an
expected homogeneous degree sequence is in fact equivalent to the
ensemble of classical random graphs. To become familiar with the
statement compare the formulas (\ref{Zer}) and (\ref{PkZ}).

Now, in order to recover the claim of equivalence between the
considered maximum-entropy models and random networks with hidden
variables one has to apply the mean field approximation to the
expression (\ref{prop0})
\begin{eqnarray}\label{prop1}
P(k_1/\theta_1)\simeq{N-1\choose k_1}\langle p_{1j}\rangle^{k_1}
(1-\langle p_{1j}\rangle)^{(N-1-k_1)},
\end{eqnarray}
where $\langle p_{1j}\rangle=\sum_jp_{1j}/(N-1)=\langle
k_{1}\rangle/(N-1)$ (\ref{Pkki}). The assumption of sparse
networks enables further simplification of the distribution
\begin{eqnarray}\label{prop2}
P(k_1/\theta_1)\simeq\frac{e^{-\langle k_1\rangle}\langle
k_1\rangle^{k_1}}{k_1!},
\end{eqnarray}
recovering the result previously derived by S\"{o}derberg
\cite{SodPRE2002} for uncorrelated networks with hidden variables.
The Poissonian propagator (\ref{prop2}) indicates the mentioned
equivalence of the considered maximum-entropy networks and the
well-known class of uncorrelated random networks with hidden
variables (Lagrange multipliers characterizing nodes correspond to
hidden attributes).

\par As pointed out in \cite{BogPRE2003}, the result is indeed very
strong since it also holds for random networks with two-point
degree correlations (see Eq. (\ref{prop3}) and also Eq. (23) in
\cite{BogPRE2003}). The key point to notice with reference to the
last expression is that in the region of small degrees, due to the
Poissonian fluctuations (c.f. (\ref{Pk_fdt2}) and (\ref{prop2})),
the {\it real} degree distributions observed in single
realizations of the considered networks strongly differ from the
desired degree distribution $P(\langle k_i\rangle)$. On the other
hand, in the limit of large degrees the real distribution and the
expected one converge. The interplay between the two distributions
has been carefully studied in our previous paper \cite{Fron2005b}.

\subsection{Networks with two-point correlations}

One can show that probability of a graph $G$ in the ensemble
(\ref{corH}) can be transformed into the same form (\ref{Pk1PG})
as the one for random networks with an expected degree sequence,
where the linkage probability is given by (\ref{corpij}).
Performing the same calculations as in the case of ensembles
analyzed in the previous subsection, one can prove that in the
limit of sparse networks the degree distribution of a specific
node is Poissonian (\ref{prop2})
\begin{equation}\label{prop3}
P(k_1/\{\Theta_{1,i}\})\simeq\frac{e^{-\langle k_1\rangle}\langle
k_1\rangle^{k_1}}{k_1!},
\end{equation}
where $\langle k_1\rangle=\sum_{i}p_{1i}$ and $p_{1i}$ represents
the connection probability given by (\ref{corpij}). Again, the
last formula supports the claimed equivalence between the analyzed
maximum-entropy networks with two-point correlations and the class
of correlated networks with hidden variables \cite{BogPRE2003}.

\section{Conclusions}

In this paper we have extended the information-theoretic approach
to random networks that was very recently proposed by Park and
Newman \cite{ParkPRE2004}. We have concentrated on fluctuations
over ensembles of undirected networks with non-interacting edges,
including random networks with a given degree sequence and
networks characterized by two-point correlations. We have studied
a few fluctuation-dissipation relations characterizing
susceptibilities of different networks to changes in the external
fields. In the case of networks with a given degree sequence, we
have argued that the scale-free topologies of real networks may
arise as a result of self-organization of real systems into sparse
structures with the low susceptibility to random external
perturbations. Finally, we have also shown that maximum-entropy
networks are equivalent to random networks with hidden variables.

\section{Acknowledgments}

A.F. acknowledges financial support from the Foundation for Polish
Science (FNP 2005). A.F. and J.A.H. were partially supported by
the State Committee for Scientific Research in Poland (Grant No.
1P03B04727). The work has also been supported by the European
Commission Project CREEN FP6-2003-NEST-Path-012864.


\begin{thebibliography}{7}
\bibitem{footnote2} One can also decide to study digraphs, weighted graphs, regular graphs and so forth.
\bibitem{footnote4} In a finite network, the cut-off of degree distribution $P(k)\sim k^{-\gamma}$ may be estimated from $\int_{k_{max}}^{\infty}P(k)dk=1/N$ yielding $k_{max}=N^{1/(1-\gamma)}.$
\end{thebibliography}
\end{document}